\documentclass[runningheads]{llncs}

\usepackage[T1]{fontenc}
\usepackage{pifont}

\usepackage{graphicx}

\usepackage{amsmath,mathtools,amssymb,amsfonts,bm}

\usepackage{subcaption}
\usepackage{graphicx}
\usepackage{caption}
\usepackage{booktabs}
\usepackage{multirow}
\usepackage{hyperref}
\hypersetup{
    colorlinks=true,
    linkcolor=blue,
    filecolor=cyan,      
    urlcolor=magenta,
    }
\usepackage[nameinlink]{cleveref}
\usepackage{color}
\usepackage[table]{xcolor}

\usepackage{wrapfig}
\usepackage{url}
\usepackage{array}

\definecolor{myorange}{rgb}{1, 0.647, 0}
\definecolor{myblue}{rgb}{.118, 0.565, 1}

\begin{document}
\title{DermoSegDiff: A Boundary-aware Segmentation Diffusion Model for Skin Lesion Delineation}
\titlerunning{DermoSegDiff: A boundary-aware segmentation diffusion}

\author{Afshin Bozorgpour\inst{1}$^\dag$ \and
Yousef Sadegheih\inst{1}$^\dag$ \and
Amirhossein Kazerouni\inst{2}$^\dag$ \and
Reza Azad \inst{3} \and
Dorit Merhof \inst{1,4}}

\authorrunning{A. Bozorgpour et al.}

\institute{
    Institute of Image Analysis and Computer Vision, Faculty of Informatics and Data Science, University of Regensburg, Regensburg, Germany \and
    School of Electrical Engineering, Iran University of Science and Technology, Iran
    \and
    Faculty of Electrical Engineering and Information Technology, RWTH Aachen University, Aachen, Germany \and
    Fraunhofer Institute for Digital Medicine MEVIS, Bremen, Germany \\ 
    \email{ \{dorit.merhof@ur.de\}} \\
    $\dag${\small \textit{ Indicates equal contribution}}
}
% % \url{http://www.springer.com/gp/computer-science/lncs} \and
% % 
\maketitle              % typeset the header of the contribution
\begin{abstract}

Skin lesion segmentation plays a critical role in the early detection and accurate diagnosis of dermatological conditions. Denoising Diffusion Probabilistic Models (DDPMs) have recently gained attention for their exceptional image-generation capabilities. Building on these advancements, we propose DermoSegDiff, a novel framework for skin lesion segmentation that incorporates boundary information during the learning process. Our approach introduces a novel loss function that prioritizes the boundaries during training, gradually reducing the significance of other regions. We also introduce a novel U-Net-based denoising network that proficiently integrates noise and semantic information inside the network. Experimental results on multiple skin segmentation datasets demonstrate the superiority of DermoSegDiff over existing CNN, transformer, and diffusion-based approaches, showcasing its effectiveness and generalization in various scenarios. The implementation is publicly accessible on \href{https://github.com/mindflow-institue/dermosegdiff}{GitHub}.

\end{abstract}
%
%
%
% ---- Introduction ----
\section{Introduction}
% ---- ------------ ----

In medical image analysis, skin lesion segmentation aims at identifying skin abnormalities or lesions from dermatological images. Dermatologists traditionally rely on visual examination and manual delineation to diagnose skin lesions, including melanoma, basal cell carcinoma, squamous cell carcinoma, and other benign or malignant growths. However, the accurate and rapid segmentation of these lesions plays a crucial role in early detection, treatment planning, and monitoring of disease progression. Automated medical image segmentation methods have garnered significant attention in recent years due to their potential to enhance diagnosis result accuracy and reliability. The success of these models in medical image segmentation tasks can be attributed to the advancements in deep learning techniques, including convolutional neural networks (CNNs) \cite{azad2022medical,ronneberger2015u,isensee2021nnu}, implicit neural representations \cite{molaei2023implicit} and vision transformers \cite{wang2021boundary,azad2023advances}.

Lately, Denoising Diffusion Probabilistic Models (DDPMs) \cite{ho2020denoising} have gained considerable interest due to their remarkable performance in the field of image generation. This newfound recognition has led to a surge in interest and exploration of DDPMs, propelled by their exceptional capabilities in generating high-quality and diverse samples. Building on this momentum, researchers have successfully proposed new medical image segmentation methods that leverage diffusion models to tackle this challenging task \cite{kazerouni2023diffusion}. EnsDiff \cite{wolleb2022diffusion} utilizes ground truth segmentation as training data and input images as priors to generate segmentation distributions, enabling the creation of uncertainty maps and an implicit ensemble of segmentations. Kim et al. \cite{kim2023diffusion} propose a novel framework for self-supervised vessel segmentation. MedSegDiff \cite{wu2022medsegdiff} introduces DPM-based medical image segmentation with dynamic conditional encoding and FF-Parser to mitigate high-frequency noise effects. MedSegDiff-V2 \cite{wu2023medsegdiff} enhances it with a conditional U-Net for improved noise-semantic feature interaction.

Boundary information has proven crucial in the segmentation of skin images, particularly when it comes to accurately localizing and distinguishing skin lesions from the surrounding healthy tissue \cite{liu2022region,wang2021boundary,kervadec2019boundary}. Boundary information provides spatial relationships between different regions within the skin and holds greater significance compared to other areas. By emphasizing these regions during the training phase, we can achieve more accurate results by encouraging the model to focus on intensifying boundary regions while reducing the impact of other areas. However, most diffusion-based segmentation methods overlook this importance and designate equal importance to all regions. Another critical consideration is the choice of a denoising architecture, which directly impacts the model's capacity to learn complex data relationships. Most methods have followed a baseline approach \cite{ho2020denoising,nichol2021improved}, neglecting the fact that incorporating semantic and noise interaction within the network more effectively.

To address these shortcomings, we propose a novel and straightforward framework called \textbf{DermoSegDiff}. Our approach tackles the abovementioned issues by considering the importance of boundary information during training and presenting a novel denoising network that facilitates a more effective understanding of the relationship between noise and semantic information. Specifically, we propose a novel loss function to prioritize the distinguishing boundaries in the segmentation. By incorporating a dynamic parameter into the loss function, we increase the emphasis on boundary regions while gradually diminishing the significance of other regions as we move further away from the boundaries. Moreover, we present a novel U-Net-based denoising network structure that enhances the integration of guidance throughout the denoising process by incorporating a carefully designed dual-path encoder. This encoder effectively combines noise and semantic information, extracting complementary and discriminative features. Our model also has a unique bottleneck incorporating linear attention \cite{shen2021efficient} and original self-attention \cite{dosovitskiy2020image} in parallel. Finally, the decoder receives the output, combined with the two outputs transferred from the encoder, and utilizes this information to estimate the amount of noise. Our experimental results demonstrate the superiority of our proposed method compared to CNN, transformer, and diffusion-based state-of-the-art (SOTA) approaches on ISIC 2018 \cite{codella2019skin}, PH$^2$ \cite{mendoncca2013ph}, and HAM10000 \cite{tschandl2018ham10000} skin segmentation datasets, showcasing the effectiveness and generalization of our method in various scenarios. Contributions of this paper are as follows: \ding{182} We highlight the importance of incorporating boundary information in skin lesion segmentation by introducing a novel loss function that encourages the model to prioritize boundary areas. \ding{183} We present a novel denoising network that significantly improves noise reduction and enhances semantic interaction, demonstrating faster convergence compared to the baseline model on the different skin lesion datasets. \ding{184} Our approach surpasses state-of-the-art methods, including CNNs, transformers, and diffusion-based techniques, across four diverse skin segmentation datasets.

\begin{figure*}[!t]
	\centering
	\begin{subfigure}{0.48\textwidth}
		\includegraphics[width=\textwidth]{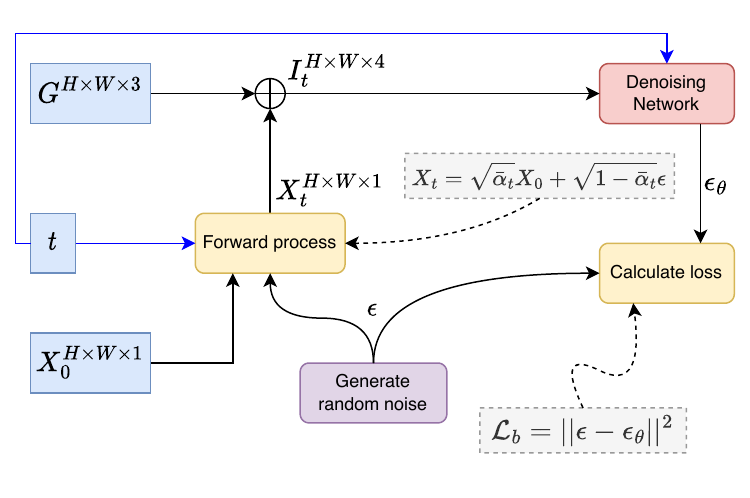}
		\caption{Baseline}
		\label{fig:baseline}
	\end{subfigure} \hfill
	\begin{subfigure}{0.48\textwidth}
		\includegraphics[width=\textwidth]{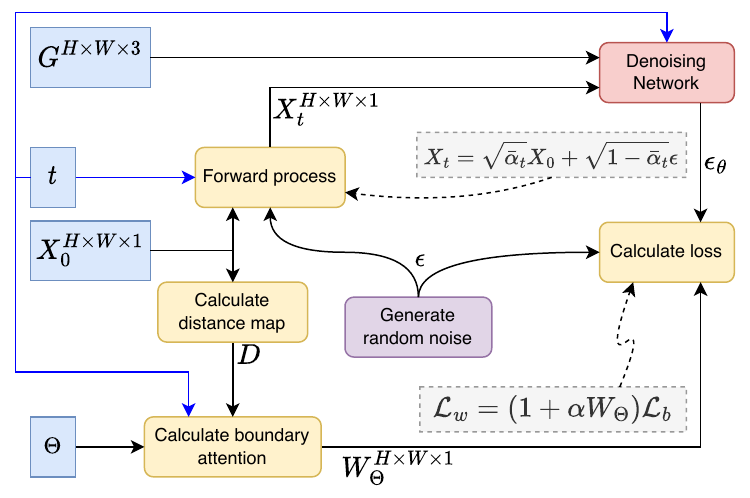}
		\caption{DermoSegDiff}
		\label{fig:pipeline}
	\end{subfigure}
	\caption{(a) illustrates the architecture of the baseline, and (b) presents our proposed DermoSegDiff framework.}
 \label{fig:method}
\end{figure*}

% ---- Methods ----
\section{Method}
\Cref{fig:method} provides an overview of our baseline DDPM model and presents our proposed \textbf{DermoSegDiff} framework for skin lesion segmentation. While traditional diffusion-based medical image segmentation methods focus on denoising the noisy segmentation mask conditioning by the input image, we propose that incorporating boundary information during the learning process can significantly improve performance. By leveraging edge information to distinguish overlapped objects, we aim to address the challenges posed by fuzzy boundaries in difficult cases and cases where lesions and backgrounds have similar colors. We begin by presenting our baseline method. Subsequently, we delve into how the inclusion of boundary information can enhance skin lesion segmentation and propose a novel approach to incorporate this information into the learning process. Finally, we introduce our network structure, which facilitates the integration of guidance through the denoising process more effectively.

\subsection{Baseline}
The core architecture employed in this paper is based on DDPMs \cite{ho2020denoising,wolleb2022diffusion} (see \Cref{fig:baseline}). Diffusion models primarily utilize $T$ timesteps to learn the underlying distribution of the training data, denoted as $q(x_0)$, by performing variational inference on a Markovian process. The framework consists of two processes: \textit{forward} and \textit{reverse}. During the forward process, the model starts with the ground truth segmentation mask ($x_0 \in \mathbb{R}^{H\times W \times 1}$) and adds a Gaussian noise in successive steps, gradually transforming it into a noisy mask:
\begin{equation}
	q\left(x_t \mid x_{t-1}\right)=\mathcal{N}\left(x_t ; \sqrt{1-\beta_t} \cdot x_{t-1}, \beta_t \cdot \mathbf{I}\right), \forall t \in\{1, \ldots, T\},
\end{equation}
in which ${\beta}_1,\dots,{\beta}_{t-1},{\beta}_T$ represent the variance schedule across diffusion steps. We can then simply sample an arbitrary step of the noisy mask conditioned on the ground truth segmentation as follows:
\begin{equation}
	q\left(\mathbf{x}_t \mid \mathbf{x}_0\right)=N\left(\mathbf{x}_t ; \sqrt{\bar{\alpha}_t} \mathbf{x}_0,\left(1-\bar{\alpha}_t\right) \mathbf{I}\right)
\end{equation}
\begin{equation}
{{x}_{t}}=\sqrt{{{{\bar{\alpha }}}_{t}}}{{x}_{0}}+\sqrt{1-{{{\bar{\alpha }}}_{t}}}\epsilon,
\end{equation}
where ${{{\alpha }}_{t}}:=1-{{\beta }_{t}}$, ${{\bar{\alpha }}_{t}}:=\prod\nolimits_{j=1}^{t}{{{\alpha }_{j}}}$ and $\epsilon \sim \mathcal{N}(0,\textbf{I})$. In the reverse process, the objective is to reconstruct the original structure of the mask perturbed during the diffusion process given the input image as guidance ($g \in \mathbb{R}^{H\times W \times 3}$), by leveraging a neural network to learn the underlying process. To achieve this, we concatenate the $x_t$ and $g$, and denote the concatenated output as $I_t:=x_t \mathbin\Vert g $, where $I_t \in \mathbb{R}^{H\times W \times (3+1)}$. Hence, the reverse process is defined as 
\begin{equation}
	p_\theta\left(\mathbf{x}_{t-1} \mid \mathbf{x}_t\right)=\mathcal{N}\left(\mathbf{x}_{t-1} ; \mu_\theta\left(I_t, t\right), \Sigma_\theta\left(I_t, t\right)\right),
\end{equation} where Ho et al. \cite{ho2020denoising} conclude that instead of directly predicting $\mu_\theta$ using the neural network, we can train a model to predict the added noise, $\epsilon_\theta$, leading to a simplified objective as $\mathcal{L}_{b}=\left\|\epsilon-\epsilon_\theta\left(I_t, t\right)\right\|^2$.
\subsection{Boundary-Aware Importance}
While diffusion models have shown promising results in medical image segmentation, there is a notable limitation in how we treat all pixels of a segmentation mask equally during training. This approach can lead to saturated results, undermining the model's performance. In the case of segmentation tasks like skin lesion segmentation, it becomes evident that boundary regions carry significantly more importance than other areas. This is because the boundaries delineate the edges and contours of objects, providing crucial spatial information that aids in distinguishing between the two classes. To address this issue, we present \textbf{DermoSegDiff}, which effectively incorporates boundary information into the learning process and encourages the model to prioritize capturing and preserving boundary details, leading to a faster convergence rate compared to the baseline method. Our approach follows a straightforward yet highly effective strategy for controlling the learning denoising process. It focuses on intensifying the significance of boundaries while gradually reducing this emphasis as we move away from the boundary region utilizing a novel loss function. As depicted in \Cref{fig:method}, our forward process aligns with our baseline, and both denoising networks produce output $\epsilon_\theta$. However, the divergence between the two becomes apparent when computing the loss function. We define our loss function as follows:
\begin{equation}
    \mathcal{L}_{w} =(1 + \alpha W_{\Theta})  \left\|\epsilon-\epsilon_\theta\left(x_t, g, t\right)\right\|^2
\end{equation}
where $W_{\Theta} \in \mathbb{R}^{H\times W \times 1}$ is a dynamic parameter intended to increase the weight of noise prediction in boundary areas while decreasing its weight as we move away from the boundaries (see \Cref{fig:impact-w}). $W_{\Theta}$ is obtained through a two-step process involving the calculation of a distance map and subsequent computation of boundary attention. Additionally, $W_{\Theta}$ is dynamically parameterized, depending on the point of time (t) at which the distance map is calculated. It means it functions as a variable that dynamically adjusts according to the specific characteristics of each image at time step $t$.

Our distance map function operates by taking the ground truth segmentation mask as input. Initially, it identifies the border pixels by assigning a value of one to them while setting all other pixels to zero. To enhance the resolution of the resulting distance map, we extend the border points horizontally from both the left and right sides by $\left\lceil H\% \right\rceil$ (e.g., for a $256\times256$ image, each row would have seven boundary pixels). To obtain the distance map, we employ the distance transform function \cite{kimmel1996sub}, which is a commonly used image processing technique for binary images. This function calculates the Euclidean distance between each non-zero (foreground) pixel in the image and the nearest zero (background) pixel. The result is a gray-level image where the intensities of points within foreground regions are modified to represent the distances to the closest boundaries from each individual point. To normalize the intensity levels of the distance map and improve its suitability as a dynamic weighting matrix $W_{\Theta}$, we employ the technique of gamma correction from image processing to calculate the boundary attention. By adjusting the gamma value, we gain control over the overall intensity of the distance map, resulting in a smoother representation that enhances its effectiveness in the loss function.

\begin{figure}[!t]
    
    \centering
    \includegraphics[width=0.92\textwidth]{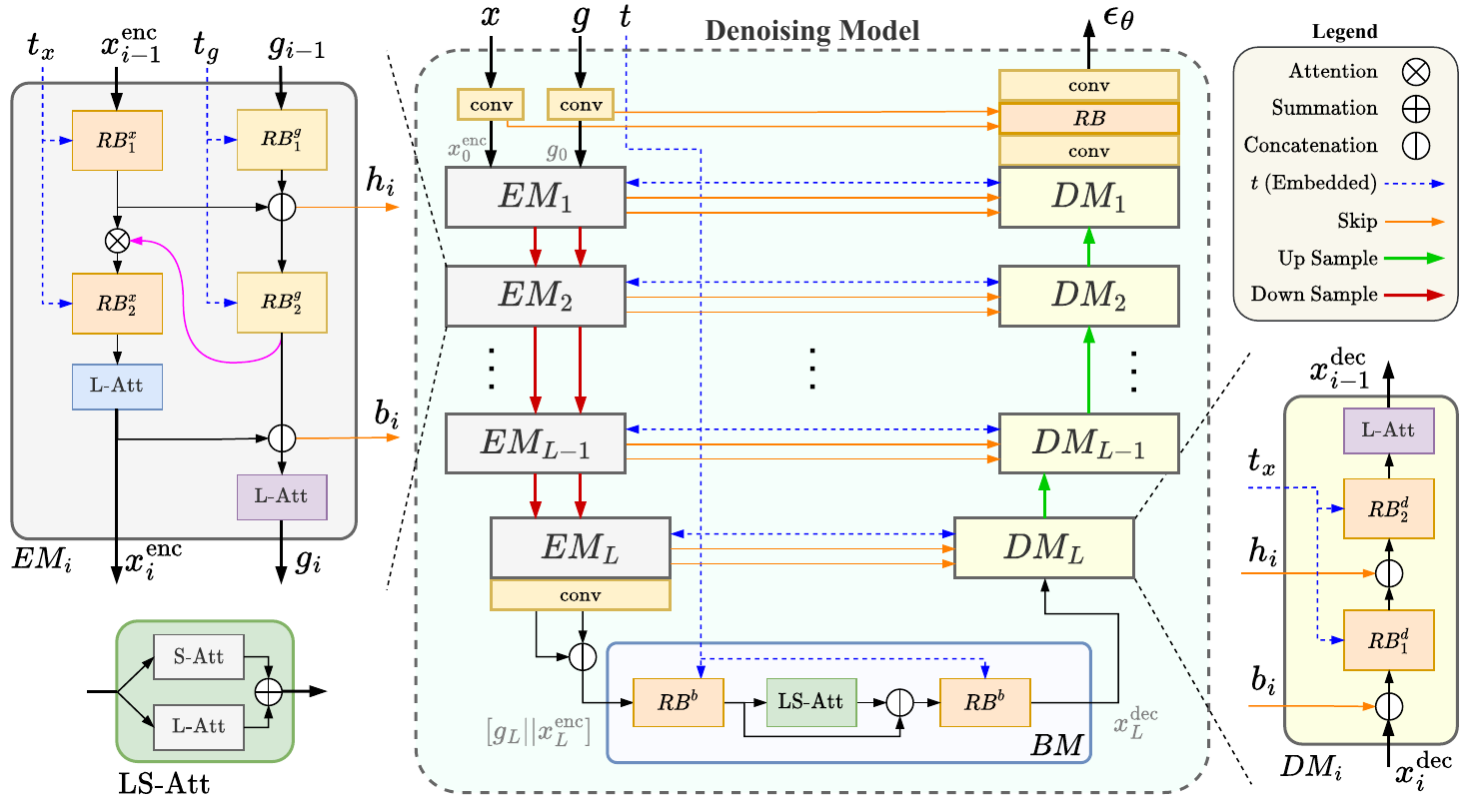}
    \caption{The overview of the proposed denoising network architecture. The notation L-Att, RB, EM, DM, LS-Att, and S-Att correspond to the Linear Attention, ResNet Block, Encoder Modules, Decoder Modules, Linear Self-Attention, and Self-Attention modules, respectively.}
    
    \label{fig:network}
\end{figure}
\subsection{Network Architecture}
\noindent\textbf{Encoder:} The overall architecture of our proposed denoising network is depicted in \Cref{fig:network}. We propose a modification to the U-Net network architecture for predicting added noise $\epsilon_\theta$ to a noisy segmentation mask $x_{i-1}^{enc}$, guided by the guidance image $g_{i-1}$ and time embedding $t$, where $i$ refers to the $i-th$ encoder. The encoder consists of a series of stacked Encoder Modules (EM), which are subsequently followed by a convolution layer to achieve a four-by-four tensor at the output of the encoder. Instead of simply concatenating $x_{i-1}^{enc}$ and $g_{i-1}$ and feeding into the network \cite{wolleb2022diffusion}, our approach enhances the conditioning process by employing a two-path feature extraction strategy in each Encoder Module (EM), focusing on the mutual effect that the noisy segmentation mask and the guidance image can have on each other. Each path includes two ResNet blocks (RB) and is followed by a Linear Attention (L-Att) \cite{shen2021efficient}, which is computationally efficient and generates non-redundant feature representation. To incorporate temporal information, time embedding is introduced into each RB. The time embedding is obtained by passing $t$ through a sinusoidal positional embedding, followed by a linear layer, a GeLU activation function, and another linear layer. We use two time embeddings, one for $g_{i-1}$ ($t_g$) and another for $x_{i-1}^{enc}$ ($t_x$), to capture the temporal aspects specific to each input. Furthermore, we leverage the knowledge captured by $RB_1^x$ by transferring and concatenating it with the guidance branch, resulting in $h_{i}$. By incorporating two paths, we capture specific representations that provide a comprehensive view of the data. The left path extracts noise-related features, while the right path focuses on semantic information. This combination enables the model to incorporate complementary and discriminative features. After applying $RB_{2}^g$, we introduce a feedback mechanism that takes a convolution of the $RB_{2}^g$ output and connects to the $RB_{2}^x$ input. This feedback allows the resultant features, which contain overall information about both the guidance and noise, to be shared with the noise path. By doing so and multiplying the feature maps, we emphasize important features while attenuating less significant ones. This multiplication operation acts as a form of attention mechanism, where the shared features guide the noise path to focus on relevant and informative regions. After the linear attention of the left path and before the right path, we provide another feature concatenation of these two paths, referred to as $b_{i}$. At the end of each EM block, we obtain four outputs: $h_{i}$ and $b_{i}$, which are used for skip connections from the encoder to the decoder, and resultant enriched $x_{i}^{enc}$ and $g_{i}$ are fed into the next EM block to continue the feature extraction process.

\noindent\textbf{Bottleneck:} Next, we concatenate the outputs, $x_L^{enc}$ and $g_L$, from the last EM block and pass them alongside the time embedding $t_x$ through a Bottleneck Module (BM), which contains a ResNet block, a Linear Self-Attention (LS-Att), and another ResNet block. LS-Att is a dual attention module that combines original Self-Attention (S-Att) for spatial relationships and L-Att for capturing semantic context in parallel, enhancing the overall feature representation. The output of BM is then fed into the decoder.

\noindent\textbf{Decoder:} The decoder consists of stacked Decoder Modules (DM) followed by a convolutional block that outputs $\epsilon_\theta$. The number of stacked DMs is the same as the number of EMs in the encoder. Unlike the EM blocks, which are dual-path modules, each DM block is a single-path module. It includes two consecutive RB blocks and one L-Att module. $b_{i}$ and $h_{i}$ from the encoder are concatenated with the feature map before and after applying $RB_1^d$, respectively. By incorporating these features, the decoder gains access to refined information from the encoder, thereby aiding in better estimating the amount of noise added during the forward process and recovering missing information during the learning process. In addition, to preserve the impact of noise during the decoding process, we implement an additional skip connection from $x$ to the final layer of the decoder. This involves concatenating the resulting feature map of the $DM_1$ with $x$ and passing them together through the last convolutional block to output the estimated noise $\epsilon_\theta$.

% ---- Dataset ----
\section{Results}
% ---- ------------ ----
The proposed method has been implemented using the PyTorch library (version 1.14.0) and has undergone training on a single NVIDIA A100 graphics processing unit (80 GB VRAM). The training procedure employs a batch size of $32$ and utilizes the Adam optimizer with a base learning rate of $0.0001$. The learning rate is decreased by a factor of $0.5$ in the event that there is no improvement in the loss function after ten epochs. In all experiments, we established $T$ as 250 and maintained the forward process variances as constants that progressively increased from $\beta_{start}=0.0004$ to $\beta_{end}=0.08$ linearly. Furthermore, in the training process, data augmentation techniques have been employed using Albumentations \cite{info11020125}, including spatial augmentation methods such as Affine and Flip transforms and CoarseDropout, as well as pixel augmentation methods such as GaussNoise and RGBShift transforms. For each dataset, the network was trained for 40000 iterations. Moreover, we set $\alpha$ empirically as 0.2. The duration of the training process was approximately 1.35 seconds per sample. Notably, in our evaluation process, we employ a sampling strategy to generate nine different segmentation masks for each image in the test set. To obtain a final segmentation result, we average these generated masks and apply a threshold of 0. The reported results in terms of performance metrics are based on this ensemble strategy.

\subsection{Datasets}
To evaluate the proposed methodology, three publicly available skin lesion segmentation datasets, ISIC 2018~\cite{codella2019skin}, PH$^2$~\cite{mendoncca2013ph}, and HAM10000~\cite{tschandl2018ham10000} are utilized. The same pre-processing criteria described in~\cite{azad2022transnorm} are used to train and evaluate the first three datasets mentioned. The HAM10000 dataset is also a subset of the ISIC archive containing 10015 dermoscopy images along with their corresponding segmentation masks. 7200 images are used as training, 1800 as validation, and 1015 as test data. Each sample of all datasets is downsized to $128\times128$ pixels using the same pre-processing as~\cite{alom2018recurrent}.

\begin{table*}[!t]
\renewcommand{\r}[1]{\textcolor{red}{#1}}
\renewcommand{\b}[1]{\textcolor{blue}{#1}}

    \centering
    \caption{Performance comparison of the proposed method against the SOTA approaches on skin lesion segmentation benchmarks. \b{Blue} indicates the best result, and \r{red} displays the second-best.}
    \label{tab:skin comparison}
    \resizebox{\textwidth}{!}{
    \begin{tabular}{l||cccc||cccc||cccc} 
    \toprule
    \multirow{2}{*}{\bf{Methods}}       &\multicolumn{4}{c||}{\bf{ISIC 2018}}    &\multicolumn{4}{c||}{$\mathbf{PH^2}$}  &\multicolumn{4}{c}{\bf{HAM10000}}\\ 
    \cline{2-13}                        & \bf{DSC} & \bf{SE} & \bf{SP} & \bf{ACC}&\bf{DSC} & \bf{SE} & \bf{SP} & \bf{ACC}&\bf{DSC} & \bf{SE} & \bf{SP} & \bf{ACC} \\ 
    \midrule
    U-Net \cite{ronneberger2015u}       &   0.8545 &   0.8800 &   0.9697 &   0.9404 &   0.8936 &   0.9125 &   0.9588 &   0.9233 &   0.9167 &   0.9085 &   0.9738 &   0.9567 
    \\
    DAGAN \cite{lei2020skin}            &   0.8807 &   0.9072 &   0.9588 &   0.9324 &   0.9201 &   0.8320 &   0.9640 &   0.9425 &   -     &   -     &   -     &   - 
    \\
    TransUNet \cite{chen2021transunet}  &   0.8499 &   0.8578 &   0.9653 &   0.9452 &   0.8840 &   0.9063 &   0.9427 &   0.9200 &   0.9353 &   0.9225 &   \b{0.9851} &   0.9649 
    \\
    %MCGU-Net \cite{}                   &    0.8950 &   0.8480 &   0.9860 &   0.9550 &   0.9263 &   0.8322 &   0.9714 &   0.9537 &   ????? &   ????? &   ????? &   ????? 
    %\\
    %MedT \cite{valanarasu2021medical}   &   0.8389 &   0.8252 &   0.9637 &   0.9358 &   0.9122 &   0.8472 &   0.9657 &   0.9416 &   ????? &   ????? &   ????? &   ????? 
    %\\
    %FAT-Net \cite{wu2022fat}            &   0.8903 &\b{0.9100}&   0.9699 &   0.9578 &   0.9440 &\b{0.9441}&   0.9741 &\b{0.9703}&   ????? &   ????? &   ????? &   ?????
    %\\
    %TMU-Net \cite{azad2022contextual}  &   {0.9059}&   0.9038 &   0.9746 &   0.9603 &   0.9414 &   0.9395 &   0.9756 &   0.9647 &   ????? &   ????? &   ????? &   ?????
    %\\
    Swin-Unet \cite{cao2021swin}        &   0.8946 &   0.9056 &   0.9798 &\b{0.9645}&   0.9449 &   0.9410 &   0.9564 &\b{0.9678}&   0.9263 &   0.9316 &   0.9723 &   0.9616 
    \\
    DeepLabv3+ \cite{chen2018encoder}   &   0.8820 &   0.8560 &   0.9770 &   0.9510 &   0.9202 &   0.8818 &\b{0.9832}&   0.9503 &   0.9251 &   0.9015 &   0.9794 &   0.9607 
    \\
    Att-UNet \cite{schlemper2019attention}                    &   0.8566 &   0.8674 &   \b{0.9863} &   0.9376 &   0.9003 &   0.9205 &   0.9640 &   0.9276 &   0.9268 &\b{0.9403}&   0.9684 &   0.9610 
    \\
    UCTransNet \cite{wang2022uctransnet}                  &   0.8838 &   \b{0.9825} &   0.8429 &   0.9527 &   0.9093 &\r{0.9698}&   0.8835 &   0.9408 &   0.9346 &   0.9205 &   0.9825 &\r{0.9684} 
    \\
    MissFormer \cite{huang2021missformer}                  &   0.8631 &   \r{0.9690} &   0.8458 &   0.9427 &   0.8550 &\b{0.9738}&   0.7817 &   0.9050 &   0.9211 &   0.9287 &   0.9725 &   0.9621 
    \\
    % Baseline \cite{wolleb2022diffusion&     0.8876 &   0.9066 &   0.9524 &   0.9371 &   0.9223 &   0.9406 &   0.9571 &   0.9462 &   0.9132 &   0.9243 &   0.9407 &   0.9452 
    Baseline (EnsDiff) \cite{wolleb2022diffusion} &   0.8775 &   0.8358 &0.9812&   0.9502 &   0.9117 &   0.8752 &   0.9774 &   0.9431 &   0.9277 &   0.9213 &   0.9771 &   0.9625 
    \\
    \midrule
    \rowcolor[HTML]{C8FFFD}
    \bf{DermoSegDiff-A}                 &   \b{0.9005} &   0.8761 &   0.9811 &   \r{0.9587} &\r{0.9450}&   0.9296 &   0.9810 &   0.9637 &\r{0.9386}&   0.9308 & 0.9814 & 0.9681 
    \\
    \rowcolor[HTML]{C8FFFD}
    \bf{DermoSegDiff-B}                 &\r{0.8966} &   0.8642 &   \r{0.9828} &   0.9575   &\b{0.9467}&   0.9308 &\r{0.9814}&\r{0.9650}&\b{0.9430}&\r{0.9326}&\r{0.9839}&\b{0.9704}
    \\
    % \midrule
    % \rowcolor[HTML]{C8FFFD}
    % \bf{DermoSegDiff-A}                 &   0.9066 &   0.8881 &   0.9811 &   0.9615 &\r{0.9450}&   0.9296 &   0.9810 &   0.9637 &\r{0.9386}&   0.9308 &\r{0.9814}&\r{0.9681} 
    % \\
    % \rowcolor[HTML]{C8FFFD}
    % \bf{DermoSegDiff-B}                 &   0.9078 &   0.8960 &   0.9792 &   0.9616 &\r{0.9450}&   0.9308 &\r{0.9814}&   0.9650 &\b{0.9430}&\b{0.9326}&\b{0.9839}&\b{0.9704}
    % \\
    % \midrule
    % \rowcolor[HTML]{C8FFFD}
    % \bf{DermoSegDiff-A}                 &   0.9078 &   0.8960 &   0.9792 &   0.9616 &\r{0.9450}&   0.9296 &   0.9810 &   0.9637 &\r{0.9386}&   0.9308 &\r{0.9814}&\r{0.9681} 
    % \\
    % \rowcolor[HTML]{C8FFFD}
    % \bf{DermoSegDiff-B}                 &   0.9091 &   0.9027 &   0.9778 &\r{0.9620}&\b{0.9467}&   0.9308 &\r{0.9814}&   0.9650 &\b{0.9430}&\b{0.9326}&\b{0.9839}&\b{0.9704}
    % \\
    \bottomrule
    \end{tabular}
    }
\end{table*}

\begin{figure}[!t]
    \centering
    \resizebox{0.9\textwidth}{!}{
        \begin{tabular}{@{} *{8}c @{}}
            \includegraphics[width=0.18\textwidth]{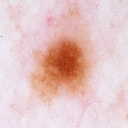} &
            \includegraphics[width=0.18\textwidth]{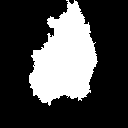} &
            \includegraphics[width=0.18\textwidth]{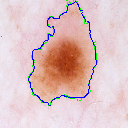} &
            \includegraphics[width=0.18\textwidth]{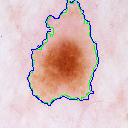} &
            \includegraphics[width=0.18\textwidth]{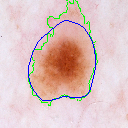} &
            \includegraphics[width=0.18\textwidth]{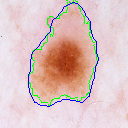} &
            \includegraphics[width=0.18\textwidth]{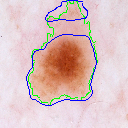} \\
            \includegraphics[width=0.18\textwidth]{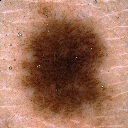} &
            \includegraphics[width=0.18\textwidth]{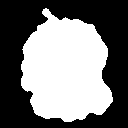} &
            \includegraphics[width=0.18\textwidth]{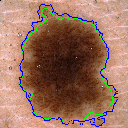} &
            \includegraphics[width=0.18\textwidth]{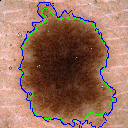} &
            \includegraphics[width=0.18\textwidth]{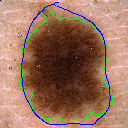} & 
            \includegraphics[width=0.18\textwidth]{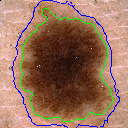} &
            \includegraphics[width=0.18\textwidth]{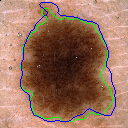}\\
            {\small {Image}} & {\small Ground Truth} & {\small DermoSegDiff-B} & {\small DermoSegDiff-A} & {TransUNet} & {\small Baseline} & {\small U-Net}
        \end{tabular}
    }
    \caption{Visual comparisons of different methods on the ISIC 2018 skin lesion dataset. Ground truth boundaries are shown
    in \textcolor{green}{green}, and predicted boundaries are shown in \textcolor{blue}{blue}.} 
    
    \label{fig:visualcomparison_isic}
\end{figure}
\subsection{Quantitative and qualitative results}
\Cref{tab:skin comparison} presents the performance analysis of our proposed DermoSegDiff on all four skin lesion segmentation datasets. The evaluation incorporates several metrics, including Dice Score (DSC), Sensitivity (SE), Specificity (SP), and Accuracy (ACC), to establish comprehensive evaluation criteria. In our notation, the model with the baseline loss function is referred to as DermoSegDiff-A, while the model with the proposed loss function is denoted as DermoSegDiff-B. The results demonstrate that DermoSegDiff-B surpasses both CNN and Transformer-based approaches, showcasing its superior performance and generalization capabilities across diverse datasets.  Specifically, our main approach demonstrates superior performance compared to pure transformer-based methods such as Swin-Unet \cite{cao2021swin}, CNN-based methods like DeepLabv3+ \cite{chen2018encoder}, and hybrid methods like UCTransNet \cite{wang2022uctransnet}. Moreover, DermoSegDiff-B exhibits enhanced performance compared to the baseline model (EnsDiff) \cite{wolleb2022diffusion}, achieving an increase of +2.18\%, +3.83\%, and +1.65\% in DSC score on ISIC 2018, PH$^2$, and HAM10000 datasets, respectively. Furthermore, in \Cref{fig:visualcomparison_isic}, we visually compare the outcomes generated by various skin lesion segmentation models. The results clearly illustrate that our proposed approach excels in capturing intricate structures and producing more accurate boundaries compared to its counterparts. This visual evidence underscores the superior performance achieved by carefully integrating boundary information into the learning process.

%%%%%%%%%%%% Ablation
\section{Ablation studies}
\begin{figure}[!t]
    \centering
    \resizebox{\textwidth}{!}{
        \begin{tabular}{@{} *{9}c @{}}
        \includegraphics[width=0.2\textwidth]{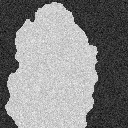} &
        \includegraphics[width=0.2\textwidth]{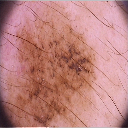} &
        \includegraphics[width=0.2\textwidth]{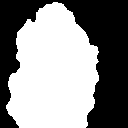} &
        \includegraphics[width=0.2\textwidth]{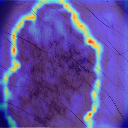} &
        \includegraphics[width=0.2\textwidth]{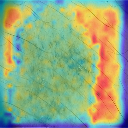} &
        \includegraphics[width=0.2\textwidth]{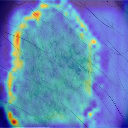} &
        \includegraphics[width=0.2\textwidth]{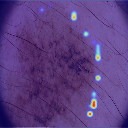} &
        \includegraphics[width=0.2\textwidth]{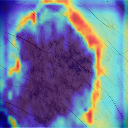} &
        \includegraphics[width=0.2\textwidth]{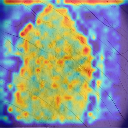} \\
        {\small Mask at $0.05T$} & {\small Guidance Image} & {\small Ground Truth} & {\small DSD-B: $x_3$} & {\small DSD-A: $x_3$} & {\small DSD-B: $g_3$} & {\small DSD-A: $g_3$}  & {\small DSD-B: feedback}  & {\small DSD-A: feedback} \\
        \\
        \includegraphics[width=0.2\textwidth]{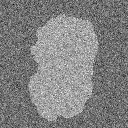} &
        \includegraphics[width=0.2\textwidth]{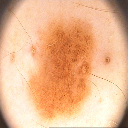} &
        \includegraphics[width=0.2\textwidth]{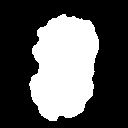} &
        \includegraphics[width=0.2\textwidth]{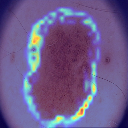} &
        \includegraphics[width=0.2\textwidth]{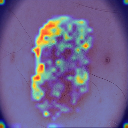} &
        \includegraphics[width=0.2\textwidth]{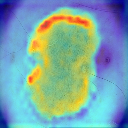} &
        \includegraphics[width=0.2\textwidth]{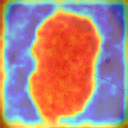} &
        \includegraphics[width=0.2\textwidth]{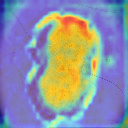} &
        \includegraphics[width=0.2\textwidth]{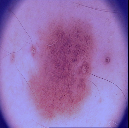} \\
        {\small Mask at $0.25T$} & {\small Guidance Image} & {\small Ground Truth} & {\small DSD-B: $x_3$} & {\small DSD-A: $x_3$} & {\small DSD-B: $g_3$} & {\small DSD-A: $g_3$}  & {\small DSD-B: feedback}  & {\small DSD-A: feedback}  
    \end{tabular}
    }
    \caption{An illustration of how our proposed loss function concentrates on the segmentation boundary in contrast to the conventional $\mathcal{L}_b$ loss in DermoSegDiff-A. The heatmaps are obtained from the $EM_3$ using GradCAM \cite{selvaraju2017grad}. Notably, DSD is an abbreviation of DermoSegDiff.} 
    
    \label{fig:boundary_isic}
\end{figure}
\begin{figure}[!t]
    \centering
    \resizebox{0.9\textwidth}{!}{
        \begin{tabular}{@{} *{6}c @{}}
        \includegraphics[width=0.17\textwidth]{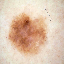} &
        \includegraphics[width=0.17\textwidth]{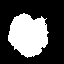} &
        \includegraphics[width=0.17\textwidth]{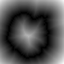} &
        \includegraphics[width=0.17\textwidth]{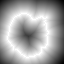} &
        \includegraphics[width=0.17\textwidth]{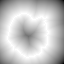} &
        \includegraphics[width=0.17\textwidth]{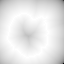} \\
        {\small Guidance Image} & {\small Ground Truth} & {\small Distance Map} & {\small $W_{\Theta}$ at 0.25T} & {\small $W_{\Theta}$ at 0.50T} & {\small $W_{\Theta}$ at 0.75T}
    \end{tabular}
    }
    
    \caption{An illustration of how the $W_{\Theta}$ variable varies dependent on the network's current time step of diffusion.} \label{fig:impact-w}
\end{figure}

\Cref{fig:boundary_isic} illustrates the effects of our innovative loss function. The heatmaps are produced utilizing the GradCAM~\cite{selvaraju2017grad}, which visually represents the gradients of the output originating from the $EM_3$. Incorporating a novel loss function results in a shift of emphasis towards the boundary region, leading to a 0.51\% enhancement compared to DermoSegDiff-A in the overall DSC score on the ISIC 2018 dataset. The analysis reveals a distinct behavior within our model. In the noise path, the model primarily emphasizes local boundary information, while in the guidance branch, it aims to capture more global information. This knowledge is then transferred through feedback to the noise branch, providing complementary information. This combination of local and global information allows our model to effectively leverage both aspects and achieve improved results. \Cref{fig:impact-w} depicts the evolution of $W_{\Theta}$ with respect to $T$. In the initial stages of the denoising process, when the effect of noise is significant, the changes in the boundary area are relatively smooth. During this phase, the model focuses on capturing more global information about the image. As the denoising process progresses and it becomes easier to distinguish between the foreground and background in the resulting image, the weight shifts, placing increased emphasis on the boundary region while disregarding the regions that are further away from it. Additionally, as we approach $x_0$, the emphasis on the boundary information becomes more pronounced. These observations highlight the adaptive nature of $W_{\Theta}$ and its role in effectively preserving boundary details during the denoising process.
\begin{figure*}[!t]
	\centering
	\begin{subfigure}{0.46\textwidth}
		\resizebox{\textwidth}{!}{
        \begin{tabular}{@{} *{3}c @{}}
        \includegraphics[width=0.2\textwidth]{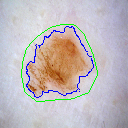} &
        \includegraphics[width=0.2\textwidth]{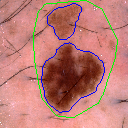} &
        \includegraphics[width=0.2\textwidth]{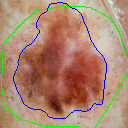}
    \end{tabular}
    }
        
		\caption{Annotation limitation}
		\label{fig:annotation}
	\end{subfigure} \hfill
	\begin{subfigure}{0.46\textwidth}
				\resizebox{\textwidth}{!}{
        \begin{tabular}{@{} *{3}c @{}}
        \includegraphics[width=0.2\textwidth]{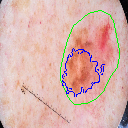} &
        \includegraphics[width=0.2\textwidth]{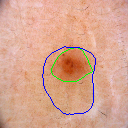} &
        \includegraphics[width=0.2\textwidth]{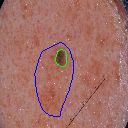}
    \end{tabular}
    }   
        
		\caption{Model limitation}
		\label{fig:model limit}
	\end{subfigure}
        
	\caption{(a) Illustrates the limitation imposed by annotation of the dataset, and (b) presents some of the limitations of our proposed model. Ground truth boundaries are shown in \textcolor{green}{green}, and predicted boundaries are shown in \textcolor{blue}{blue}. }
 \label{fig:limitation}
\end{figure*}

% ---- Conclusion ----
\section{Limitations}
% ---- ------------ ----

Despite these promising results, there are also some limitations. For example, some annotations within the datasets may not be entirely precise. \Cref{fig:annotation} portrays certain inconsistencies in the annotations of data. However, despite these annotation challenges, our proposed method demonstrates superior precision in the segmentation of skin lesions in comparison to the annotators. The results indicate that with more meticulous annotation of the masks, our proposed approach could have achieved even higher scores across all evaluation metrics. It is worth noting that there were instances where our model deviated from the accurate annotation and erroneously partitioned the area. \Cref{fig:model limit} depicts instances where our proposed methodology fails to segment the skin lesion accurately. The difficulty in accurately demarcating the boundary between the foreground and background in skin images arises from the high similarity between these regions and requires more work that we aim to address in future work.

% ---- Conclusion ----
\section{Conclusion}
% ---- ------------ ----
This paper introduced the \textbf{DermoSegDiff} diffusion network for skin lesion segmentation. Our approach introduced a novel loss function that emphasizes the importance of the segmentation's boundary region and assigns it higher weight during training. Further, we proposed a denoising network that effectively models the noise-semantic information and results in performance improvement.

\bibliographystyle{splncs04}
\bibliography{15.bib}




% LNCS Homepage, \url{http://www.springer.com/lncs}. Last accessed 4
% \end{thebibliography}
\end{document}